\documentclass[aps,twocolumn,showpacs,groupedaddress,superscriptaddress,amsmath,amssymb]{revtex4}

\usepackage{graphicx}
\usepackage{dcolumn}
\usepackage{bm}

\begin{document}

\title{Scaling Properties of Flexible Membranes from Atomistic Simulations: \\
       Application to Graphene}

\author{J.\ H.\ Los}
\affiliation{Institute for Molecules and Materials, Radboud University 
Nijmegen, Heyendaalseweg 135, 6525 AJ Nijmegen, The Netherlands}
\author{M.\ I.\ Katsnelson}
\affiliation{Institute for Molecules and Materials, Radboud University 
Nijmegen, Heyendaalseweg 135, 6525 AJ Nijmegen, The Netherlands}
\author{O. V. Yazyev}
\affiliation{Ecole Polytechnique F\'ed\'erale de Lausanne (EPFL),
Institute of Theoretical Physics, CH-1015 Lausanne, Switzerland}
\affiliation{Institut Romand de Recherche Num\'erique en Physique
des Mat\'eriaux (IRRMA), CH-1015 Lausanne, Switzerland}
\author{K.\ V.\ Zakharchenko}
\affiliation{Institute for Molecules and Materials, Radboud University 
Nijmegen, Heyendaalseweg 135, 6525 AJ Nijmegen, The Netherlands}
\author{A.\ Fasolino}
\affiliation{Institute for Molecules and Materials, Radboud University 
Nijmegen, Heyendaalseweg 135, 6525 AJ Nijmegen, The Netherlands}

\date{\today}

\begin{abstract}
Structure and thermodynamics of crystalline membranes are characterized 
by the long wavelength behavior of the normal-normal correlation function 
$G(q)$. We calculate $G(q)$ by Monte Carlo and Molecular Dynamics 
simulations for a quasi-harmonic model potential and for a realistic 
potential for graphene. To access the long wavelength limit 
for finite-size systems (up to 40000 atoms) we introduce a Monte Carlo 
sampling based on collective atomic moves (wave moves). 
We find a power-law behaviour 
$G(q)\propto q^{-2+\eta}$ with the same exponent $\eta \approx 0.85$ 
for both potentials. This finding supports, from the 
microscopic side, the adequacy of the scaling theory of membranes 
in the continuum medium approach, 
even for an extremely rigid material like graphene.
\end{abstract}

\maketitle

Collective phenomena involving infinitely many degrees of freedom are 
often characterized by scaling laws with power-law behavior of correlation 
functions. In three dimensional systems, this behavior occurs 
only at critical points \cite{Wilson,Binney,Yeomans}. In two dimensions
(2D) the situation is different, and a whole temperature interval with
``almost broken symmetry'' and power-law decay of 
correlation functions frequently appears, the Kosterlitz-Thouless (KT) 
transition in 2D superfluids and superconductors \cite{KT} being a prototype example. 
Existence of real long range order, where correlation functions remain 
non-zero in the limit of infinite distance, is forbidden in such cases by the 
Mermin-Wagner theorem \cite{mermin} due to the divergence of the 
contribution of soft modes to relevant thermodynamic properties. 
The theory of flexible membranes \cite{nelson} embedded in higher dimensions
is an important part of the statistical mechanics of 2D systems. 
Here, we investigate the scaling behavior of 
crystalline flexible membranes by means of atomistic simulations, using
graphene \cite{first,r1,r2}, the simplest known membrane, as an example. 

In the flat phase, the membrane in-plane and out-of-plane displacements 
are parametrized by a $D$-component `stretching' phonon field 
$u_\alpha({\bf x})$, $\alpha=1...D$, and by a $d_c=d-D$ component 
out-of-plane height fluctuation $h({\bf x})$, where $d$ is the 
space dimension and $D$ is the membrane dimension.  Softening of 
bending modes makes this situation very similar to the KT model. 
A minimal phenomenological model for membranes is 
just the elasticity theory described by the Hamiltonian \cite{Peliti,nelson}:
\begin{equation} \label{Free_F_1}
H = \frac{1}{2}\int\! d^D\!x\,(\kappa (\nabla^2 h)^2 + 
\mu u_{\alpha\beta}^2+\frac{\lambda}{2}u_{\alpha\alpha}^2)
\end{equation}
where $\kappa$, $\mu$ and $\lambda$ are bending rigidity, 
shear modulus and Lam\'{e} coefficient and 
\begin{equation}\label{deftens}
u_{\alpha\beta} = \frac{1}{2}(\partial_{\alpha}u_{\beta}+
\partial_{\beta}u_{\alpha}+
\partial_{\alpha}\!h\,\partial_{\beta}\!h)
\end{equation}
is the strain tensor. 
In harmonic approximation, by neglecting the last, non-linear,
term in Eq.~(\ref{deftens}), the bending ($h$)
and stretching (${\bf u}$) modes are decoupled.

The Hamiltonian (\ref{Free_F_1})  is quadratic in the phonon degrees of freedom ${\bf u}$ which can be eliminated by Gaussian integration \cite{nelson,Peliti}. In this way, the Hamiltonian can be rewritten only in terms of the Fourier components of the height  $h$  as the sum of a harmonic bending energy, quadratic in $h$,  and an anharmonic energy, quartic in $h$, that results from  the coupling of bending and stretching modes \cite{Peliti}.
If one neglects the latter term, the membrane becomes crumpled
at any finite temperature with, for $D=2$, the mean square  height fluctuations
$\langle h^2 \rangle \sim L^2$ and normal-normal correlation functions that diverge logarithmically
at large distances. Nelson and Peliti \cite{Peliti} suggested that the
above anharmonic term stabilizes the flat phase at least at temperatures much
smaller than $\kappa$. This flat phase  is described by an effective bending
rigidity $\kappa(q)\sim q^{-\eta}$ and effective elastic moduli with power-law
dependencies on $q$ that partially suppress long wavelength bending fluctuations. As
a result, the normal-normal correlation function remains finite, although 
$\langle h^2 \rangle$ still diverges as $\langle h^2 \rangle \sim L^{2\zeta} $ with 
$\zeta=1-\eta/2$ \cite{nelson}. Thus, the flat phase is not
truly flat, but still exhibits rather strong corrugation.
The model (\ref{Free_F_1}), which is called the model of phantom membranes, has 
a transition to a crumpled phase at temperature of the order of $\kappa$. The long wavelength limit was solved within the Self Consistent Screening Approximation in 
Ref.~\onlinecite{Doussal} yielding $\eta=0.821$.
The discretized version of this model was investigated by Bowick {\it et al.} by 
means of Monte Carlo simulations giving $\eta\approx 0.72$ \cite{Bowick}.
The term `phantom' means that the model does not include self-avoidance, the natural condition 
of true physical systems. It is assumed that self-avoidance 
removes the phase transition to  the high temperature crumpled phase while the 
scaling properties of the `flat' phase remain the same as in phantom membranes. 
However, any kind of accurate statement about the model can be justified only 
in the limit $d_c\rightarrow\infty$ and, strictly speaking, nothing can be said 
rigorously for the real case of  $d=3$, $D=2$ and $d_c=1$. 

To characterize the long 
wavelength limit of the height fluctuations  we compare the results of atomistic simulations to the predictions of this theory for the normal-normal correlation functions  $G(q)=\langle|{\bf n}_q|^2\rangle$. Starting from Eq.~(\ref{Free_F_1}) an expression for $G(q)$ has been given  from general scaling considerations\cite{Peliti,nelson,Fas2007} in the form of an effective Dyson equation
\begin{equation}
G_a^{-1}\left( q\right) =G_0^{-1}\left( q\right) +\Sigma \left( q\right)
\label{Gq}
\end{equation}
where $G_0$ is the value derived in harmonic approximation
\begin{equation}
G_0\left( q\right) =\frac {T N}{\kappa S_0 q^2}
\label{G0}
\end{equation}
and the self energy is
\begin{equation}
\Sigma \left( q\right) =\frac {A S_0}{N} q^2\left( \frac{q_0}{q}\right) ^\eta
\label{Sigma}
\end{equation}
with $N$ the number of atoms, $S_0=L_xL_y/N$ the area per atom, 
$T$ the temperature in units of energy, $q_0=2\pi \sqrt{B/\kappa }$, $B$ the two-dimensional
bulk modulus \cite{note} and $A$ 
an unknown numerical factor. 

Until recently, this phenomenological continuum model was the only way to describe  
the statistical mechanics of membranes since all known real membranes~\cite{nelson} 
were too complicated for atomistic models. The situation has been changed drastically by 
the discovery of graphene~\cite{first} which is the first example of a truly 
two-dimensional system (just one atom thick) and, thus, a prototype crystalline 
membrane~\cite{r1,r2}. The experimental observation of ripples 
in freely hanged graphene~\cite{jannik} 
stimulated a large theoretical activity
\cite{Fas2007,castroneto,GKV,guinea,wehling,zakhar,foster}. In particular, 
using the accurate bond order potential for carbon LCBOPII~\cite{los1}, 
we were able to simulate structural and thermodynamical  properties of graphene 
at finite temperatures~\cite{Fas2007,zakhar} by straightforward Monte Carlo (MC) 
simulations. The simulations confirmed the existence of thermally induced 
intrinsic ripples at finite temperatures resulting in strong anharmonic effects. 
However, we found that the normal-normal correlation function could not be described by Eq.~(\ref{Gq}) over the whole range of $q$~\cite{Fas2007}. 
In fact, $G(q)$ followed
the power law resulting from the harmonic approximation (phonon picture, $\eta=0$) at 
large enough $q$, but, after bending, 
 at smaller $q$'s we found a drop of the correlation 
functions not compatible with a power law. Our conjecture at that time was 
that the extreme rigidity of graphene could be the reason why it could not 
be described by the phenomenological theory of membranes in a continuum medium 
approach~\cite{Peliti}. However, we felt that this point deserved further investigation. 
Here, we focus on the low-$q$ region in order to establish firmly whether a 
scaling law exists and, if so, to determine the
scaling exponent. To this purpose, we simulate large systems, introduce new 
MC moves for  phase space sampling and examine more than one model of the 
interatomic forces, including a simple quasi-harmonic (QH) model that yields 
a not too rigid membrane and the extremely rigid case of graphene which is 
well described by LCBOPII. In addition, for the QH model we verified 
ergodicity of our MC simulations by comparing with Molecular Dynamics 
(MD) results. 

We begin by considering a relatively simple QH model 
with energy given by:
\begin{eqnarray*}
U = \frac{1}{2} \sum_i \sum_{j \neq i} \left( K_r  ( r_{ij} - r_{eq} )^2 +
K_{\theta} \sum_{k\neq i,j}( y_{ijk} - y_{eq} )^2 \right)
\end{eqnarray*}
where the summations over $j$ and $k$ are over the nearest-neighbors of atom $i$,
$ y_{ijk} = \cos{\theta_{ijk}} $ and $ y_{eq} = \cos{\theta_{eq}} $,
with $ r_{eq} = 1.42 $~\AA\ and $ \theta_{eq} =2 \pi/3 $ the ground state
equilibrium nearest neighbor distance and bond angle in graphene. The
stretching and the bending force constants, $ K_r = 22 $~eV~\AA$^{-2}$ and 
$ K_{\theta} = 4 $~eV, respectively, were chosen to yield  
elastic moduli  for isotropic and  uniaxial compressions  equal to those for the LCBOPII~\cite{zakhar}.

\begin{figure}
\includegraphics[clip,width=0.9\linewidth]{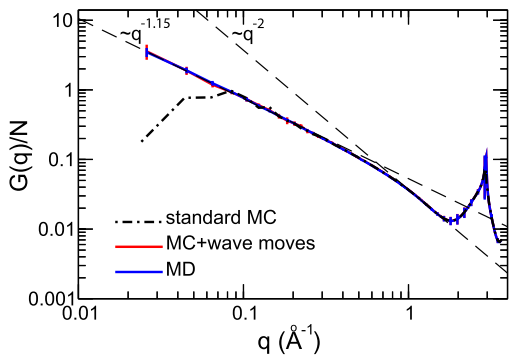}
\caption{ (color online) Normal-normal correlation functions $G(q)/N$ 
calculated for a graphene system with  
$N  = 37888$  by ordinary MC simulations (red-dashed line), 
MD simulations and MC simulations with wave moves with the QH potential. 
The dashed lines show the asymptotic harmonic behavior with power laws $q^{-2}$ for large $q$ and the long-wavelength limit $q^{-(2-\eta)}$ with $\eta=0.85$. } 
\label{fig1}
\end{figure}

In Fig.~\ref{fig1} we show the function $G(q)/N$ (dotted line) calculated by extensive 
standard Monte Carlo simulations in the canonical ensemble at 300 K for a 
system with $N=37888$, $L_x=314.82$~\AA\ and $L_y=315.24$~\AA\ and periodic boundary 
conditions in the 
$xy$-plane. Starting from the Bragg peak at $q=4\pi/(3r_{eq})=2.94$~\AA$^{-1}$ and going towards lower $q$ 
we find, first, the  power law $\eta=0$ due to the harmonic contribution, then, 
a smaller slope followed by a drop at the smallest $q<0.08$~\AA$^{-1}$ which
corresponds to a wavelength of about 75~\AA. This drop is similar to the one mentioned above and found 
previously in Ref.~\onlinecite{Fas2007} with the LCBOPII for graphene. 
These results are obtained by averaging over many configurations in the canonical 
ensemble obtained by the ordinary MC procedure which is based on random 
displacements of randomly chosen individual atoms
and volume (area) fluctuations with a Metropolis acceptance
rule. By using Eq.~(\ref{G0}) we find that the bending constant for the QH potential 
is $\kappa=0.4$ eV, much softer than the 1.1~eV appropriate for graphene \cite{Fas2007}, due to 
neglected interactions beyond first neighbors. 
The observation that 
also the simple QH model shows a suppression
of long wavelength excitations made us think of the possibility that
standard MC is not an efficient sampling technique in this
case. To resolve this issue we (i) extended our MC phase space sampling with a new type
of collective trial events that we call `wave moves' described below, and (ii)
performed MD simulations for the
QH model~\cite{note2}, allowing a direct comparison with the
MC results, with and without wave moves.
The equivalence of time averages in MD simulations with ensemble averages in 
MC simulations  guarantees that the system is in thermodynamical equilibrium (ergodic).

In Fig.~\ref{fig1} we compare the results of standard MC with the results obtained by MD 
and by MC with the addition of wave moves. The MD results coincide with the standard 
MC in the range where the latter is described by a power law, but does not 
show the drop at small $q$ and keeps the same slope till the smallest 
possible $q$ allowed by our finite size system. The results of 
MC simulations with wave moves coincide for all $q$'s with those obtained by MD, 
implying that the system is in thermodynamic equilibrium. Both curves 
display a power-law behavior for the whole range of $q$ in the 
long wavelength limit. A best fit of the data yields an exponent $\eta=0.85$. 

\begin{figure}
\includegraphics[clip,width=0.9\linewidth]{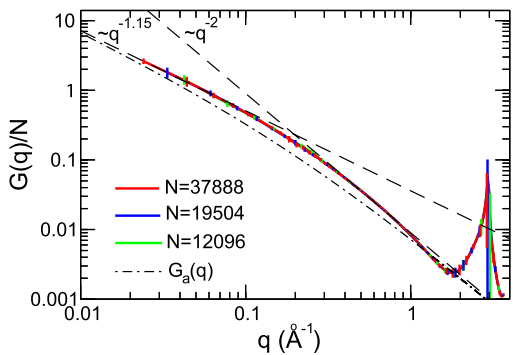}
\caption{ (color online) Normal-normal correlation functions $G(q)/N$ 
calculated for three systems with  
$N= 12096$ ($L_x = 177.08$~\AA, $ L_y = 178.92$~\AA), 
$N  = 19504$ ($L_x = 226.27$~\AA, $ L_y = 225.78$~\AA),
$N  = 37888$ ($L_x = 314.82$~\AA, $ L_y = 315.24$~\AA) by 
MC simulations with wave moves with LCBOPII. The dashed lines 
show the asymptotic harmonic behavior with power laws $q^{-2}$ 
for large $q$ and the long-wavelength limit $q^{-(2-\eta)}$ with 
$\eta=0.85$. The dashed-dotted line is $G_a$ of Eq.~(\ref{Gq}) 
with the coefficients fixed by the asymptotic behavior. 
One can see that the crossover is much sharper in the simulations.} 
\label{fig2}
\end{figure} 

A wave move consists of a transversal, wavelike 
displacement of all atoms in the $z$-direction, perpendicular 
to the graphene plane. For a given wavevector $ {\bf q} $ there
are two possible, linearly independent wave excitations,
yielding $z$-coordinate displacements for all atoms $i$ 
\begin{eqnarray*}
\Delta {\bf z}_i &=& (0.5 - R) A_{S,q} \cos( {\bf q} {\bf r}_i )~~~\mbox{and}\\
\Delta {\bf z}_i &=& (0.5 - R) A_{S,q} \sin( {\bf q} {\bf r}_i )
\end{eqnarray*}
where $ {\bf r}_i $ is the 3D position of atom $i$ and
$ R $ is a random number between 0 and 1. The amplitude
$ A_{S,q} $ is chosen such that the acceptance rate
for such a wave move is between 0.4 and 0.5.
The appropriate value of $ A_{S,q} $ depends on the size
of the 2D box $ S=L_x L_y $ and on the wavevector $ {\bf q} $
(see below).

Due to the periodic boundary conditions in the $x$- and
$y$-directions the candidate wavevectors for wave moves can be restricted to a
set on a 2D grid:
\begin{eqnarray*}
{\bf q} = \left(m_x \frac{2 \pi}{L_x} , m_y \frac{2 \pi}{L_y},0\right)
\end{eqnarray*}
with integer $ m_x $ and $ m_y $. This set was further bounded by applying
only wave moves of long wavelengths since short
wavelengths are already efficiently sampled
by the individual atom displacement trials. 
Hence, we consider  a finite set of $ (m_x, m_y) $-pairs corresponding to
$q$-vectors within a circular region with
radius $ q_{max} $ around $ {\bf q} = 0 $. This set was kept
constant during the entire simulation. We choose  $ q_{max} $
 equal to the $q$-value below which  $ G(q) $
starts to bend down in standard MC simulation. More precisely,
we took $ q_{max} \simeq 0.16 $, corresponding to a minimal 
wavelength of 40~\AA. 
Since transversal phonon modes have quadratic dispersion $ \omega (q) \sim q^2 $, 
the energy change associated with a wave move 
behaves as $ \Delta E_{wm} \sim A_{S,q}^2 q^2 $. Therefore, we
took $ A_{S,q} = A_S/q $ to obtain similar acceptance rates for 
each of the allowed $q$-vectors, as was indeed
confirmed by our simulations. This choice leaves 
one adjustable parameter, $A_S$. For different system sizes,
the appropriate $A_S$ roughly scales as $ A/S $, but a
correction is required to fine-tune the acceptance rate.
On average, a wave move was attempted every MC step by choosing randomly 
one of the 2$N_q$ possible waves. Here, $ N_q $ is the number of allowed $q$-vectors
(or $ (m_x, m_y) $-pairs) and the factor 2 comes from the fact
that each wavevector yields two possible waves: a sine and a
cosine wave. Another random number $ R \in (0,1)$ was then pulled
to fix the amplitude $ (0.5-R)A_S/q $. Following the Metropolis procedure,
a wave move is always accepted if the energy change
$ \Delta E_{wm} $ is negative,
whereas for $ \Delta E_{wm} > 0 $ it is accepted with
probability $ P=\exp(-\beta \Delta E_{wm}) $, requiring another
random number $ R' \in (0,1) $ to decide for acceptance when
$ R' \leq P $ or rejection when $ R' > P $.

\begin{figure}
\includegraphics[clip,width=0.9\linewidth]{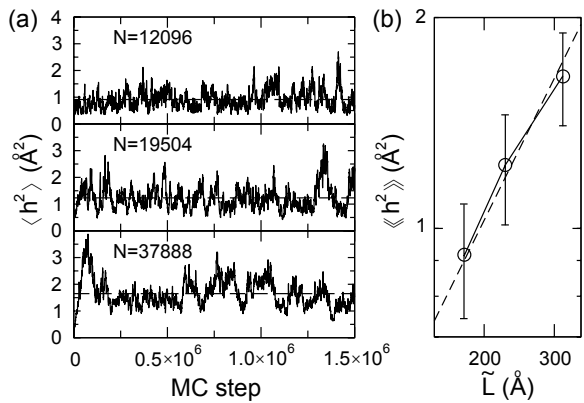}
\caption { 
(a) Quadratic out-of-plane displacement $\langle h^2 \rangle$,  
averaged over all particles, as a function of the MC step for
the same systems as in Fig.~\ref{fig2}. The dashed horizontal lines 
denote $\langle \langle h^2 \rangle \rangle$, the average 
$\langle h^2 \rangle$ over all MC steps except the first 
$3\times10^5$ steps of equilibration. 
(b) $\langle \langle h^2 \rangle \rangle$ as a function of the average
linear system size $\tilde{L}=\sqrt{L_xL_y}$ compared to the scaling 
law $\langle \langle h^2 \rangle \rangle=C L^{2-\eta}$ with $C$=0.00232
and $\eta=0.85$ (dashed line). Both axis are in logarithmic scale.}
\label{fig3}
\end{figure}

MD simulations are much more demanding than MC simulations and 
are not within reach for the rather complex LCBOPII potential for the 
present system size. The previous results with the QH harmonic potential, however, show
that equilibrium
can be reached using MC with wave moves. The correlation function $G(q)$ calculated 
by MC with wave moves
for LCBOPII are shown in Fig.~\ref{fig2} for three system sizes.
Again, we see the crossover from the harmonic behavior 
to a power law with $\eta=0.85$ up to the smallest wavevectors. 
The main difference with the results obtained with the QH potential 
is that, due to a higher bending rigidity, the crossover between 
the two power laws is shifted to lower $q$ values. 
Moreover, we note that for $q>1$~\AA$^{-1}$ there is a deviation from a power 
law behavior just before the Bragg peak. 

Finally, in Fig.~\ref{fig3}(a) we show the average out-of-plane displacement 
$\langle h^2 \rangle$ corresponding to the simulations for LCBOPII of Fig.~\ref{fig2} which shows 
large fluctuations. In Fig.~\ref{fig3}(b) we plot the values of $\langle h^2 \rangle$ 
averaged over all MC steps 
as a function of the system size in comparison with the expected 
scaling law $\exp(2-\eta)$.  Although it would have been impossible 
to deduce the scaling exponent from the three points in Fig.~\ref{fig3}(b) 
due to the large error originating from the large fluctuations, 
these results are certainly compatible with the scaling exponent 
$\eta$ found by a fit of $G(q)$. With  $\langle h^2 \rangle=1.65$ \AA$^2$ for  $\tilde{L}=315$ \AA~ and $\eta=0.85$ we estimate $\sqrt{\langle h^2 \rangle}\approx $ 9 \AA~ for $L$=1$\mu$m, well in the range of measured values~\cite{jannik}.

In summary, we have shown by atomistic simulations that, in thermodynamic 
equilibrium, crystalline membranes display a power-law scaling behavior 
of the normal-normal correlation function, in qualitative agreement 
with continuum medium theory. For different models 
of interactions with different rigidities, we found the same exponent of 
anomalous bending rigidity $\eta\approx 0.85$. We have demonstrated that 
the efficiency of MC simulations for this type of systems can be greatly 
improved by introducing collective wave moves. On the basis of our results, 
we conclude that despite its extreme rigidity, graphene behaves as a prototype 
membrane opening new ways to study the intriguing physics of membranes on 
a system with well known interatomic interactions.

This work is part of the research program of the `Stichting voor Fundamenteel Onderzoek der Materie (FOM)', which is financially supported by the `Nederlandse Organisatie voor Wetenschappelijk
Onderzoek (NWO)'.


\begin{thebibliography}{99}

\bibitem{Wilson} K.\ G.\ Wilson and J.\ Kogut, Phys.\ Rep. {\bf 12}, 75 (1974).

\bibitem{Binney} J.\ J.\ Binney, N.\ J.\ Dowrick, A.\ J.\ Fisher, and M.\ E.\ J.\ Newman, {\it The Theory of Critical Phenomena} (Oxford University Press, Oxford, 1992).

\bibitem{Yeomans} J.\ M.\ Yeomans, {\it Statistical Mechanics of Phase Transitions} (Oxford University Press, Oxford, 1992).

\bibitem{KT} J.\ M.\ Kosterlitz and D.\ J.\ Thouless, J.\ Phys.\ C {\bf 6}, 1181 (1973).  

\bibitem{mermin} N.\ D. Mermin and H.\ Wagner, Phys.\ Rev.\ Lett. {\bf 17}, 1133 (1966).

\bibitem{nelson} {\it Statistical Mechanics of Membranes and Surfaces}, edited by D.\ R.\ Nelson, T.\ Piran, and S.\ Weinberg (World Scientific, Singapore, 2004).

\bibitem{first} K.\ S.\ Novoselov, A.\ K.\ Geim, S.\ V.\ Morozov, D.\ Jiang, Y.\ Zhang, S.\ V.\ Dubonos, I.\ V.\ Grigorieva, and A.\ A.\ Firsov, Science {\bf 306}, 666 (2004).

\bibitem{r1} A.\ K.\ Geim and K.\ S.\ Novoselov, Nat.\ Mater. {\bf 6}, 183 (2007).

\bibitem{r2} M.\ I.\ Katsnelson, Mater.\ Today {\bf 10}, 20 (2007).

\bibitem{Peliti} D.\ R.\ Nelson and L.\ Peliti, J.\ Phys.\ (Paris) {\bf 48}, 1085 (1987).

\bibitem{Doussal} P.\ Le\ Doussal and L.\ Radzihovsky,  Phys.\ Rev.\ Lett. {\bf 69}, 1209 (1992).

\bibitem{Bowick} M.\ J.\ Bowick, S.\ M.\ Catterall, M.\ Falcioni, G.\ Thorleifsson, and K.\ N.\ Anagnostopoulos, J.\ Phys.\ I\ (Paris) {\bf 6}, 1321 (1996).

\bibitem {note} The $T=0$ value of $B$ for graphene according to LCBOPII has been incorrectly reported in Ref.~\onlinecite{Fas2007}. The correct value is $B =12.7$~eV/\AA$^2$.

\bibitem{jannik} J.\ C.\ Meyer, A.\ K.\ Geim, M.\ I.\ Katsnelson, K.\ S.\ Novoselov, T.\ J.\ Booth, and S.\ Roth, Nature {\bf 446}, 60 (2007).

\bibitem{Fas2007} A.\ Fasolino, J.\ H.\ Los, and M.\ I.\ Katsnelson, Nat.\ Mater. {\bf 6}, 858 (2007).

\bibitem{GKV} F.\ Guinea, M.\ I.\ Katsnelson, and M.\ A.\ H.\ Vozmediano, Phys.\ Rev.\ B {\bf 77}, 075422 (2008).

\bibitem{foster} M.\ S.\ Foster and I.L.\ Aleiner, Phys.\ Rev.\ B {\bf 77}, 195413 (2008).

\bibitem{guinea} F.\ Guinea, B.\ Horovitz, and P.\ Le\ Doussal, Phys.\ Rev.\ B {\bf 77}, 205421 (2008).

\bibitem{wehling} T.\ O.\ Wehling, A.\ V.\ Balatsky, A.\ M.\ Tsvelick, M.\ I.\ Katsnelson, and A.\ I.\ Lichtenstein, Europhys.\ Lett. {\bf 84}, 17003 (2008).

\bibitem{castroneto} E.-A.\ Kim and A.\ H.\ Castro\ Neto, Europhys.\ Lett. {\bf 84}, 57007 (2008).

\bibitem{zakhar} K.\ V.\ Zakharchenko, M.\ I.\ Katsnelson, and A.\ Fasolino, Phys.\ Rev.\ Lett.\ {\bf 102}, 046808 (2009).

\bibitem{los1} J.\ H.\ Los, L.\ M.\ Ghiringhelli, E.\ J.\ Meijer, and A.\ Fasolino, Phys.\ Rev.\ B  {\bf 72}, 214102 (2005).

\bibitem{note2} The MD simulations are obtained for a slightly 
different QH potential where the term 
$K_{\theta}  ( \cos{\theta_{ijk}} - \cos{\theta_{eq}} )^2 $
is changed to 
$\tilde{K}_{\theta} ( \theta_{ijk} - \theta_{eq} )^2 $. 
By choosing $ \tilde{K}_{\theta} = 3K_{\theta}/4 $ the two expressions 
coincide to the lowest order.

\end{thebibliography}
\end{document}